\documentclass[fleqn,12pt,twoside]{article}
\usepackage{espcrc1}

\usepackage{epsfig}
\usepackage[figuresright]{rotating}

\newcommand{\AmS}{{\protect\the\textfont2
  A\kern-.1667em\lower.5ex\hbox{M}\kern-.125emS}}

\title{Power counting in relativistic baryon chiral perturbation theory}

\author{T. Fuchs,\address[MCSD]{Institut f\"{u}r Kernphysik, Johannes
Gutenberg-Universit\"{a}t
D-55099 Mainz, Germany}%
\thanks{Supported by the Deutsche Forschungsgemeinschaft (SFB 443).}
        J. Gegelia,\addressmark\thanks{Supported by the Alexander von Humboldt
        Foundation.}
        G. Japaridze,\address{CTSPS, Clark Atlanta University, Atlanta, 
Georgia 30314, USA}
        and
S. Scherer\addressmark[MCSD]\thanks{Supported by the Deutsche
        Forschungsgemeinschaft (SFB 443).}}

\begin{document}
\maketitle

\begin{abstract}
 We consider a renormalization scheme for relativistic baryon
chiral perturbation theory which provides a simple and consistent
power counting for renormalized diagrams. As an application we
discuss the chiral expansion of the nucleon mass.
\end{abstract}

\medskip
\medskip
 The effective field theory of
(pseudo) Goldstone bosons \cite{Weinberg:1979kz,Gasser:1984yg} was
extended by Gasser, Sainio,
and \v{S}varc to include also processes involving one external 
nucleon  \cite{Gasser:1988rb} (for a recent review see, e.g.,
\cite{Scherer:2002tk}).
   One of the findings in their scheme was that higher-loop diagrams can
contribute to terms as low as ${\cal O}(q^2)$, where $q$
generically denotes a small expansion parameter such as, e.g., the
pion mass. This problem has been solved in the heavy-baryon
formulation of ChPT \cite{Jenkins:1990jv}.
   Although this approach leads to a straightforward power counting,
its disadvantage is that, in some cases, it does not provide the
correct analytic behavior even in the threshold regime
\cite{Bernard:1996cc}.
   Several methods have been suggested to reconcile power counting with the
constraints of analyticity in the relativistic approach
\cite{Ellis:1997kc,Becher:1999he,Gegelia:1999gf,Gegelia:1999qt,%
Lutz:1999yr,Fuchs:2003qc}.


The most general effective Lagrangian includes all possible
interaction terms which are compatible with the underlying
symmetries and thus provides us with all the required
counterterms.
   Since the finite parts of the counterterms are arbitrary, one
has the freedom of choosing a suitable renormalization condition.
   In this work we choose the finite parts of the counterterms so that their
contributions precisely cancel those parts of the loop diagrams
which violate the power counting. This leads us to a simple and
consistent power counting for the renormalized diagrams of a
relativistic approach \cite{Fuchs:2003qc}. As an example we
consider the nucleon self energy.



We use the standard power counting of Ref. \cite{Ecker:1995gg} together with
the Lagrangian of Ref.\ \cite{Fettes:2000gb}.

\begin{figure}[ht]
\vspace{0.1cm} \epsfxsize=14cm \centerline{\epsffile{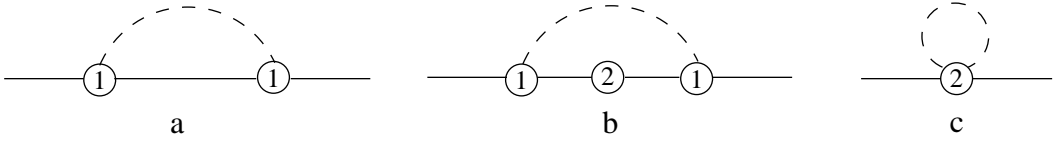}}
\vspace{0.1cm} \caption[] {\label{mnloops:fig} One-loop
contributions to the nucleon self-energy at ${\cal O}(q^4)$.}
\end{figure}

At ${\cal O}(q^4)$, the self energy  receives contact
contributions $\Sigma_{\rm contact}$ from ${\cal L}_{\pi N}^{(2)}$
and ${\cal L}_{\pi N}^{(4)}$ as well as the one-loop contributions
of Fig.\ \ref{mnloops:fig},

\begin{equation}
\Sigma=\Sigma_{\rm contact}+\Sigma_a+\Sigma_b+\Sigma_c, \ \ \
\Sigma_{\rm contact}=-4 M^2 c_1^0-2 M^4
(8e_{38}^0+e_{115}^0+e_{116}^0),
\end{equation}
\begin{eqnarray}
\label{sigmaa} \Sigma_a&=&-\frac{3
{\stackrel{\circ}{g_{A}}}_0^2}{4 F_0^2}\left\{
\vphantom{-\frac{(p^2-m^2)p\hspace{-.5em}/}{2p^2}}
(p\hspace{-.45em}/\hspace{.1em}+m)I_N
+M^2(p\hspace{-.45em}/\hspace{.1em}+m)I_{N\pi}(-p,0)\right.\nonumber\\
&&\left. -\frac{(p^2-m^2)p\hspace{-.45em}/\hspace{.1em}}{2p^2}[
(p^2-m^2+M^2)I_{N\pi}(-p,0)+I_N-I_\pi]\right\}, \\ \label{sigmab}
\Sigma_b&=&-4M^2 c_1^0\,\frac{\partial\Sigma_a}{\partial m}, \ \ \
\ \Sigma_c = 3\frac{M^2}{F_0^2}\left(2 c_1^0-c_3^0 -
\frac{p^2}{m^2}\frac{c_2^0}{n}\right) I_\pi,
\end{eqnarray}

\noindent where the expressions for the loop integrals read

$$ \label{INpi} I_{N\pi}(p,0)=2\bar\lambda-{1\over 16\pi^2}+{
\sqrt{4 m^2 M^2-\left(p^2-m^2-M^2\right)^2}\over 16\pi^2
p^2}\arccos\left( \frac{m^2+M^2-p^2}{2 m M}\right) $$
\begin{equation}
+{p^2-m^2+M^2\over 16\pi^2 p^2} \ln\left(\frac{M}{m}\right),
\end{equation}

\begin{equation}
\label{Ipi}
I_\pi=2M^2\bar{\lambda}+\frac{M^2}{8\pi^2}\ln\left(\frac{M}{m}\right),\
I_N=2m^2\bar{\lambda},\  \bar\lambda ={m^{n-4}\over (4\pi
)^2}\left\{ {1\over n-4}- {1\over 2}\left[ \ln (4\pi )+\Gamma
'(1)+1\right]\right\}. \label{IN}
\end{equation}

   The renormalization of the loop diagrams is performed in two steps.
   First we render the diagrams finite by applying the subtraction scheme
used by Gasser and Leutwyler \cite{Gasser:1984yg,Gasser:1988rb}
which we denote by modified minimal subtraction scheme of ChPT
($\widetilde{\rm MS}$)
amounting to dropping the
terms proportional to $\bar{\lambda}$ in the loop integrals.
   We choose the renormalization parameter $\mu =m$.
 In a second step, a given $\widetilde{\rm MS}$-renormalized
diagram is written as the sum of a subtracted diagram which
satisfies the power counting and a remainder which violates the
power counting and thus needs to be subtracted.
   We expand the couplings of the $\widetilde{\rm MS}$
scheme in terms of the couplings of our extended on-mass-shell (EOMS)
scheme, thus generating finite counterterms responsible for
additional finite subtractions.
   These counterterms are fixed so that the net result of combining
the counterterm diagrams with those parts of the $\widetilde{\rm
MS}$-renormalized diagrams which violate the power counting are of
the same order as the subtracted diagram.
Hence the sum of an $\widetilde{\rm MS}$-renormalized diagram and
the corresponding counterterm diagram satisfies the power
counting.

   For the case at hand, we determine the subtraction terms 
by first expanding the integrands and
coefficients in Eqs.\ (\ref{sigmaa}) and (\ref{sigmab}) in powers of
$M^2$, $p\hspace{-.45em}/\hspace{.1em}-m$, and $p^2-m^2$.
   In this expansion we keep all the terms having a chiral order which is
smaller than what is suggested by the power counting for the given
diagram.
   We then obtain
\begin{equation}
\label{sesubtrterm} \Sigma_{r,a+b+c}^{\rm subtr}=
\frac{3{\stackrel{\circ}{g_A}}_r^2}{32\pi^2 F_r^2}
\left[mM^2-\frac{(p^2-m^2)^2}{4m}\right] +\frac{3 c_1^r
{\stackrel{\circ}{g_A}}_r^2 M^2}{8 \pi^2 F_r^2}
\left[m(p\hspace{-.45em}/\hspace{.1em}+m)-\frac{3}{2}(p^2-m^2)\right].
\end{equation}
   We fix the corresponding counterterms so that they exactly cancel the
expression given by Eq.\ (\ref{sesubtrterm}).
   Finally, the renormalized self-energy expression is obtained by subtracting
Eq.\  (\ref{sesubtrterm}) from the  $\widetilde{\rm MS}$-subtracted 
versions of Eqs.\ (\ref{sigmaa}) and (\ref{sigmab}) and
replacing the $\widetilde{\rm MS}$-renormalized couplings with the
ones of our EOMS scheme.

 The physical nucleon mass at order ${\cal O}(q^4)$ can be written as 
\cite{Leutwyler:1999mz,Sainio:2001bq}
\begin{equation}
\label{massnucleonparameterization} m_N=m+k_1 M^2+k_2 M^3+k_3
M^4\ln\left(\frac{M}{m}\right)+k_4 M^4+ {\cal O}(M^5).
\end{equation}
In the EOMS renormalization scheme the coefficients $k_i$ are given by
\begin{eqnarray}
\label{parki} k_1&=&-4 c_1, \ \ \ k_2=-\frac{3
{\stackrel{\circ}{g_A}}^2}{32\pi F^2}, \ \ \ k_3=\frac{3}{32\pi^2
F^2}\left(8c_1-c_2-4 c_3
-\frac{{\stackrel{\circ}{g_A}}^2}{m}\right), \nonumber\\
k_4&=&\frac{3 {\stackrel{\circ}{g_A}}^2}{32 \pi^2 F^2 m}(1+4 c_1
m) +\frac{3}{128\pi^2F^2}c_2-16 e_{38}-2 e_{115}-2e_{116}.
\end{eqnarray}
   Comparing with Ref.\ \cite{Becher:1999he}, we see that the lowest-order
correction and those terms which are nonanalytic in the quark
mass $\hat{m}$ coincide, but the analytic $k_4$ term is different.
   This is due to the usage of a different renormalization scheme
and hence the difference between the two results is compensated by
different values of the renormalized parameters.


It is straightforward to use our approach in an iterative
procedure to renormalize higher-order loop diagrams.
   Finally, our renormalization scheme is neither restricted to the
single-nucleon sector nor to the interaction of Goldstone bosons
with fermions \cite{Fuchs:2003sh}.

\end{document}